# "A Vehicle of Symbols and Nothing More"

# George Romanes, Theory of Mind, Information, and Samuel Butler[1]

**Donald R. Forsdyke**

Queen's University, Canada

## Abstract

Today's "theory of mind" (ToM) concept is rooted in the distinction of nineteenth century philosopher William Clifford between "objects" that can be directly perceived, and "ejects," such as the mind of another person, which are inferred from one's subjective knowledge of one's own mind. A founder, with Charles Darwin, of the discipline of comparative psychology, George Romanes considered the minds of animals as ejects, an idea that could be generalized to "society as eject" and, ultimately, "the world as an eject" – mind in the universe. Yet, Romanes and Clifford only vaguely connected mind with the abstraction we call "information," which needs "a vehicle of symbols" – a material transporting medium. However, Samuel Butler was able to address, in informational terms depleted of theological trappings, both organic evolution and mind in the universe. This view harmonizes with insights arising from modern DNA research, the relative immortality of "selfish" genes, and some startling recent developments in brain research.

## Keywords

Brain, DNA as information, ejective inference, information flow, subjectivity

## Introduction

Charles Darwin's work impacted three major areas – evolution in general, neuroscience/psychology, and philosophy/theology. The demands of the first occupying much of his time, it was a relief to pass the burden to someone who shared his enthusiasm, like himself was independently wealthy and, most importantly, was four decades younger. Two quotations capture this: 'How glad I am you are so young' marked Darwin's first meeting with George John Romanes in 1874 (Forsdyke, 2001: 204). And when the German evolutionist Ernst Haeckel dismissed part of Darwin's work as an 'airy nothing,' Darwin in 1876 quipped that Romanes would 'some day, I hope, convert an "airy nothing" into a substantial theory' (Schwartz, 2010: 138). Furthermore, the young man, having trained in neurophysiology at Cambridge, and having a deep interest in both psychology and philosophy/theology, was well prepared to address the latter two areas of Darwin's work.

Thus, apart from his immediate family, during the last eight years of Darwin's life (1809-1882) it is likely that no one had more opportunity to discuss evolution with him than Romanes (1848-1894). Indeed, with permission, his mentor's unpublished manuscripts on brain evolution were incorporated by Romanes into a major work – *Mental Evolution in Animals* (Romanes, 1884; Stauffer, 1975: 463-527). In 1886 *The Times* of London hailed Romanes as 'the biological investigator upon whom the mantle of Darwin has most conspicuously descended' (Forsdyke, 2001: 220-222).

But their warm relationship was shattered at Darwin's death. To cope with his grief, and to feel a way towards deeper theological issues that resisted clear articulation, Romanes turned to poetry – a passion that, together with his philosophical/theological writings, he had either kept from the public, or had contributed anonymously (Romanes, 1889; Turner, 1974; Pleins, 2014). To his friends, he was 'the philosopher,' and he sometimes signed his letters as such. His friends included the philosopher William Clifford.

In his 1885 Cambridge Rede Lecture, Romanes addressed from a neurobiological perspective the perceived confrontation between evolution and theology engendered by Darwin's work – a confrontation that endures. But, while the lecture and associated essays (Romanes, 1882, 1885, 1886) opened up new territory, there was something missing. Assisted by advances in DNA research, we can today recognize that he – like most, but not all, of his Victorian contemporaries – did not see that the issues might become clearer if addressed in informational terms. The Victorian exception was Samuel Butler who, with Prague neurophysiologist Ewald Hering, introduced information concepts to biology in the 1870s that would provide foundations for an understanding of both brains (mental memory) and genomes (genetic memory).

This paper is in four parts. The first covers Clifford's idea of ejective inference and Romanes' theoretical exploration of its scope. The second deals with Hering-Butler information concepts. The third, extending the seminal studies of Schacter (2001) and Otis (1994), covers the controversy surrounding Butler's work. Finally, this history is related to some modern perspectives on science and theology.

## 1. Ejective inference

### *Theory of mind as ejective inference*

There is much current interest in 'theory of mind,' (ToM; Premack and Woodruff, 1978; Frith, 2012), a topic dating back to Romanes (Obiols and Berrios, 2009), who built on earlier work of Clifford. Your knowledge of mind begins with your subjective understanding of your own thought processes. To this you attach the word 'mind.' This is the seemingly inescapable frame of reference through which you view both the world and the minds of others (Tononi and Koch, 2014). But the way you view the world differs from the way you view other minds. Since I share numerous attributes with you (e.g. we belong to the same species) it would seem reasonable to suppose that my thought processes operate in the same way as yours. So you can *imagine* yourself 'in my shoes.' Based on this ToM, you can make predictions about my behaviour that will generally be correct (e.g. you would help an old lady across the road, and so would I). But there remains uncertainty. For example, many in Western societies find it difficult to conceive mindsets that lead to self-immolation for what are deemed 'terrorist' causes. Thus, my mind – not directly accessible to you – is an *inferred* entity. As such it is, in the word of Clifford, an 'eject' – not an 'object' that can be observed by your senses, even if assisted by technology (e.g. the microscope and perhaps even modern brain-scanning devices; but see Carrington and Bailey, 2009). Thus a ToM, relating to 'a subjective order in many respects analogous to my own,' can be referred to as an ejective hypothesis (Clifford, 1878: 275). So there are three 'ivities:' *subjectivity, ejectivity* and *objectivity*, of which the first is one's own, introspective, personal frame of reference.

### Mind as emergent property

Clifford, a Professor of Applied Mathematics at University College, London, thought that, despite being perceivable by others only ejectively, your mind (your subjectivity), like their minds (their subjectivities), would obey the laws of physics and chemistry that applied to the objective world. Such laws dealt with the two fundamental entities - matter and motion, the latter often displaced today by the term 'energy.' These two – matter, and the motion of that matter – composed what he called 'mind-stuff,' something with the *potential* to become constituted into a form with the property we call 'mind' (i.e. stuff that, at some level of complexity, could form mind). Thus, 'a moving molecule of inorganic matter does not possess mind or consciousness; but it possesses a small piece of mind-stuff' (Clifford, 1878: 284). Assuming mind as part of your body, then 'the facts do very strongly lead us to regard our bodies as merely complicated examples of practically universal physical rules, and their motions as determined in the same way as those of the sun and the sea' (Clifford, 1878: 279). So, 'when matter takes on the complex form of a living human brain, the corresponding mind-stuff takes the form of a human consciousness, having intelligence and volition' (Clifford, 1878: 284). Mind, circumscribed within an organism, was a property that could emerge when its mind-stuff had attained a sufficient level of complexity. Thus: 'Reason, intelligence, and volition are properties of a complex which is made up of elements themselves not rational, not intelligent, not conscious' (Clifford, 1878: 286).

### Romanes expanded the scope of ejective inference

Clifford (1878: 277) held that 'recognition of a kindred consciousness in one's fellow-beings – is clearly a condition of gregarious action among animals so highly developed as to be called conscious at all.' Romanes took this further in his three essays (Romanes, 1882, 1885, 1886). Could we ejectively infer what an animal was thinking and so predict its behaviour? Conversely, could an animal ejectively infer what we think and so predict our behaviour? While experimental studies date the emergence of human ejective powers during childhood (Obiols and Berrios, 2009; Frith, 2012), the design of such experiments in animals has proved problematic (Penn and Povinelli, 2007; Radick, 2007: 70-83). No less problematic are ejective predictions concerning what appear to be higher levels of complexity than either animals or ourselves.

Romanes (1882: 885) considered Clifford's 'essay … the most closely reasoned and profound of all his philosophical writings.' Like Clifford, he located mind within the brain (Romanes, 1882: 871):

> 'Within experience, mind is found in constant and definite association with that highly complex and peculiar disposition of living matter called a living brain, the size and elaboration of which throughout the animal kingdom stands in conspicuous proportion to the degree of intelligence displayed, and the impairment of which … entails corresponding impairment of mental processes.'

Indeed, in his Rede lecture he offered proofs that 'the grey matter of the cerebral hemispheres is the exclusive seat of mind' (Romanes, 1885: 77). But he disagreed with Clifford when it came to the scope of ejective thought being limited only to individual natural forms. 'Because within the limits of human experience mind is only known as associated with brain, it clearly does not follow that mind cannot exist in any other mode' (Romanes 1885: 92). Building on Clifford's 'kindred consciousness,' he pointed to society – the 'social organism' – that 'although composed of innumerable individual personalities, is, with regard to each of its constituent units, a part of the objective world – just as the human brain would be, were each of its constituent cells of a construction sufficiently complex to yield a separate personality' (Romanes, 1886: 51-52). And it was possible to discern in this social organism some form of subjectivity:

'Seeing that ideas are often, as it is said, "in the air" before they are condensed in the mind of individual genius, we habitually speak of the "Zeitgeist" as a kind of collective psychology, which is something other than the mere sum of all the individual minds of a generation. That is to say, we regard Society as an eject … . The ejective existence thus ascribed to Society serves as a stepping-stone to the yet more vague and general ascription of such existence to the Cosmos.'

### The world as eject

Romanes proposed that the subjective individual can extrapolate ejectively from other individuals he/she knows, to the collective of individuals we deem as a society and, ultimately, to the inanimate world around us – universal mind. Yet, according to Romanes in his paper – 'The world as an eject' – Clifford would exclude the possibility of such a cosmic eject 'unless we can show in the disposition of the heavenly bodies some morphological resemblance to the structure of the human brain.' Romanes (1886: 49-50) thought otherwise:

'For aught that we can know to the contrary, not merely the highly specialized structure of the human brain, but even that of nervous matter in general, may only be one of a thousand possible ways in which the material and dynamic conditions required for the apparition of self-consciousness can be secured. To imagine that the human brain of necessity exhausts these possibilities is in the last degree absurd. … To take the particular conditions under which alone subjectivity is known to occur upon a single planet as exhausting the possibilities of its occurrence elsewhere, is too flagrant a use of the method of simple enumeration to admit of a moment's countenance. … Therefore, … the one exhibition of subjectivity furnished to human experience, is the proof thus furnished that subjectivity is possible under *some* conditions. … The macrocosm does furnish amply sufficient opportunity … for the presence of subjectivity, even if it be assumed that subjectivity can only be yielded by an order of complexity analogous to that of a nervous system.'

What form might this cosmic subjectivity take? In the scheme of things, it is likely that you regard your mind as at a high level – much higher than, say, a stone, or even your arm muscle.

If there is something higher than your mind, it is an abstraction that you can imagine, if at all, only by extrapolation from mind. Considering 'that the world … is at least as susceptible of an ejective as it is of a objective interpretation' (Romanes, 1886: 59), then Romanes supposed that, 'just as a mathematician is able to deal symbolically with space in $n$ dimensions, while only able to conceive of space as limited to three dimensions, so I feel that I ought not to limit the abstract possibilities of mental being by what I may term the accidental conditions of my own being' (Romanes, 1886: 54). Indeed, borrowing words from the theological domain,[1] concerning his 'world eject' Romanes (1885: 92) held:

> 'There is a mode of mind which is not restricted to brain, but co-extensive with motion, is con-substantial and co-eternal with all that was, and is, and is to come; have we not at least a suggestion, that high as the heavens are above the earth, so high above our thoughts may be the thoughts of such a mind as this?'

## 2. Information

### *Information flows from medium to medium*

Like most of their Victorian contemporaries, Clifford and Romanes never quite connected mind or body to the abstract concept of 'information.' While not using this word, Clifford (1878: 279-280) recognized that different media could convey the same information. A given item of information could exist in more than one medium:

> 'A spoken sentence and the same sentence written, are two utterly unlike things, but each of them consists of elements; the spoken sentence of the elementary sounds of the language, the written sentence of its alphabet. … Or as we should say in mathematics, a sentence spoken is the same function of the elementary sounds as the same sentence written is of the corresponding letters.'

And, rather than mental 'information for,' he wrote of a mental 'representation of,' an external object – a candlestick (Clifford, 1878: 285):

'The cerebral image is an imperfect representation of the candlestick, corresponding to it point for point in a certain way. Both the candlestick and the cerebral image are matter; but one material complex *represents* the other material complex in an imperfect way.'

Concerning mind, Romanes (1886: 57) instead of 'information' used abstractions such as 'inward meaning,' 'spiritual grace,' and a reality that is 'psychic':

'Already we are able to perceive the immense significance of being able to regard any sequence of natural causation as … the merely outward manifestation of an inward meaning. Thus, for example, I am listening to a sonata of Beethoven's played by Madame Schumann. … The great reality is the mind of Beethoven communicating to my mind through the complex intervention of three different brains with their neuromuscular systems, and an endless variety of aërial vibrations proceeding from the pianoforte. The method of communication has nothing more to do with the reality communicated than have the paper and ink of this essay to do with the ideas which they serve to convey. In each case a vehicle of symbols [medium] is necessary in order that one mind should communicate with another; but in both cases this is a vehicle of *symbols*, and nothing more. Everywhere, therefore, the reality may be psychical, and the physical symbolic; everywhere matter in motion may be the outward and visible sign of an inward and spiritual grace.'

And concerning bodies, when pondering the relationship between 'germ-plasm' contained in the egg and the resulting organism, Romanes (1893) again used Clifford's 'representation:'

'The theory … supposes merely that every part of the future organism is represented in the egg-cell by corresponding material particles. And this, as far as I can understand, is exactly what the [Weismann] theory of germ-plasm supposes; only it calls the particles "molecules," and seemingly attaches more importance to … variations in their arrangement or "constitution," whatever these vague expressions may be intended to signify.'

However, prior to this work, two others – the Prague neurophysiologist Ewald Hering (1834-1918) and the ex-sheep-farmer, artist, and satirist, Samuel Butler (1835-1902) – had already begun to address in informational terms, the evolution of minds and bodies in the universe (Schacter, 2001: 106-119; Forsdyke: 2006; 2009; 2011a: 7-26). Fundamental to this was the idea that items of information, abstract in themselves, *could not exist* without media that would preserve and transmit them. Nevertheless, the 'medium is the message' (McLuhan, 1964: 23-35) is only true when 'message' refers to the preserving/transporting agency, but not when it refers to the abstract item of information contained *within* that agency. This distinction was later emphasized by the influential American biologist George Williams (1926-2010): 'Information can exist only as a material pattern, but the same information can be recorded in a variety of patterns in many different kinds of material. A message is always coded in some medium, but the medium really is not the message' (Williams, 1992: 10-13).

### Thoughts as information-flows

Your mind manifests itself to you as information-flows that you call 'thoughts.' In writing the previous sentence, I converted a thought-information-flow into a textual-information-flow that you can read and then, hopefully, understand (generate further meaningful thought-information-flows). 'If I could think to you without words, you would understand me better' lamented a friend of Samuel Butler, probably Charles Paine Pauli (Butler, 1878: 84). Instead, Pauli had to convert his thought-information-flow into a textual-information-flow or speech-information-flow. Then Butler, having received this form of information, had, in turn, to convert it back into a thought-information-flow. In an 1887 lecture he observed (Butler 1923: 190):

'The thought is not in the alphabet, nor in the words into which the letters of the alphabet are grouped. These are vehicles of thought, and in so far as they are this, they have made thought more easy, more convenient, more tidy, and have infinitely extended it; but they are not thought. Thought, and the appliances for facilitating and extending thought, grow up together as supply and demand invariably do; supply stimulates demand, and demand stimulates supply; but neither is the other. So thought stimulates language, and language, thought; feeling stimulates the nervous system, and the nervous system, feeling; but the thought lies deeper than speech, and feeling, than the nervous system.'

### Thought language preceded spoken language

The word 'language,' as pointed out by Butler in an essay dated 1890 and 1894 (Butler, 1904: 181), derives from the Latin word for tongue. This tends to make 'language' synonymous with 'spoken language,' rather than generically referring to any form of coded information transfer – a transfer from medium to medium according to some convention (code). Yet, guided by our modern understanding of protein synthesis (see later), it follows that thought-information-flow *must* have a language (or languages) of its own, that we could refer to as 'mentalese' (Pinker, 1994: 78-81). Thus, information flowed from Pauli to Butler when Pauli's mentalese was converted into spoken words, and these words, in turn, were then converted into Butler's mentalese. In his 1887 lecture Butler (1923: 200, 202) observed:

> 'As the word has only an artificial connection with the idea for which it passes current, so the idea itself has nothing but an arbitrary artificial and conventional connection with the object which it serves to bring before the mind. … Why do Englishmen and Frenchmen attach the same ideas to many external objects but not to all, and why do they differ so much more widely in language than in ideas? The reason is because ideas are far earlier things, and are, in fact, an infinitely older language, than the words which now do duty for them. The lower animals have no articulated language, but few will say that they have no ideas. A cat, for example, has very sufficient and definite ideas of a mouse or of milk. … Ideas, then, are much older than even the oldest and least specialized means of formulating and conveying them; and it is because they are so much older that they are so much more settled and differ so infinitely less in different nations. … It is only because language is so much newer than either ideas or feelings, that we are able with comparative ease to disinter its past and forecast its future; and if it were as old, it would be as mysterious as feeling is.'

As with other languages, mentalese should have syntactic structure. If mentalese evolved before spoken languages, then this pre-existing mentalese could have influenced the evolution of those languages, which hence might share common syntactical features. This viewpoint would appear to differ from Chomsky (1965), who holds that constraints underlying a 'universal grammar' arose somatically from 'physical or developmental limitations' that had 'little or nothing to do with communication' *per se*. In other words, language is an exploitation, an 'exaptation,' of something that pre-existed and served some other, essentially non-communicative, role in brain function (Fitch, 2013).

### *Perceptual distinctions are arbitrary*

Clifford (1878: 285-286) distinguished the *cerebral image* of an object from the ultimate *mental image* of that object, but held that both were contained in 'mind-stuff.' While the distinction intrigued Francis Crick in his 'scientific search for the soul' (1995: 255-263), it is not of primary concern here. But we can note that, once information in a cerebral image has been assimilated as a thought, by definition, it has become a mental image. For example, the redness of something may be conveyed in cerebral image mentalese. The evolutionary 'decision' that the mentalese conveying information on this particular wavelength of light, would be perceived as 'red' – i.e. its qualia is red – may be no different from the decision Darwin's parents made that the gurgling infant from whom information flowed to their senses, would be perceived as 'Charles.' He had to be called something. He had to be distinguished by some name. Likewise, if it were evolutionarily advantageous to distinguish colours, each wavelength would have been 'assigned' by natural selection an appropriate perceptual 'name' (colour). And if there were no advantage in associating a particular colour with a particular wavelength, then the assignment would have been random. But even if arbitrary, there *had* to be an assignment. The wavelength *had* to be 'called' something. The mechanistic basis by which 'mind-stuff' achieves this still eludes us. In 1887 the arbitrariness of the assignment of a feeling as 'pain' was noted by Butler (1923: 198-199):

'The feeling of which we are aware on being pricked by a pin has no primordial inherent inevitable connection with our being pricked, either by a pin or by anything else, but it is an arbitrary convention, invented and adopted for the purpose of enabling us more certainly and easily to recognize the combination of things and circumstances with which we have associated it, in exactly the same way as the word stone is a conventional arrangement of certain sounds, … by means of which the idea of a stone is more rapidly and certainly conveyed, though the word has no essential inherent connection with, nor resemblance to, a stone itself.'

## 3. The controversial role of Samuel Butler

### *Genes and instincts as information*

Erwin Schrödinger's *What is Life?* (1944) inspired those who gave us DNA structure and associated informational concepts in the 'classical period' of molecular biology (1945-1966; Chargaff, 1978: 85; Olby, 2009: 65-66; Forsdyke, 2011a: 18; Cobb, 2013; Battail, 2014: 148-151). Indeed, in *Mind and Matter* he touched on many of the issues raised here (Schrödinger, 1958). But where did this famous physicist (born 1887 in Vienna) get these ideas? Although there were various mid-twentieth century influences (Sloan and Fogel, 2011: 1-35), in essays dated 1925 and 1960 – published as *My View of the World* – Schrödinger (1964) repeatedly cited the *Mneme* books of Richard Semon, who had studied with Haeckel in Jena. English translations appeared in the 1920s (Schacter, 2001). We can now trace the path of fundamental informational ideas – oscillating between German and English – from Hering/Butler to Semon (1904), to Schrödinger, and then on to Francis Crick and others.

From the outset Butler saw the inheritance of parental characters by offspring in informational terms and the word 'information' was used in appropriate context (Butler 1880: 252):

'Does the offspring act as if it remembered? The answer to this question is not only that it does so act, but that it is not possible to account for either its development or its early instinctive actions upon any other hypothesis that that of its remembering, and remembering exceedingly well. The only alternative is to declare … that a living being may display a vast and varied *information* concerning all manner of details, and be able to perform most intricate operations, independently of experience and

practice.' [My italics]

Traces of this view are in his satirical *Erewhon* (Butler, 1872). When Darwin's son Francis told Butler that Hering (1870) had been thinking similarly (Forsdyke, 2006), Butler set about learning German, and a translation of Hering was in the third of his major works on evolution – *Unconscious Memory* (Butler, 1880). Butler held that the information passed from generation to generation, be it related to bodily function, or to brain function (instincts), was a 'hereditary memory.' He used 'memory,' not as a process of recall (e.g. 'she's got a good memory'), but as *stored information* – a usage familiar today as computer memory. Butler's view was opposed by Romanes, who pointed out that Charles Kingsley (1867) had described bird migration likewise:

> 'Yon woodwren … essayed the British Channel, and was blown back, … [yet] he felt, nevertheless, that "that was water he must cross," he knew not why: but something told him that his mother had done it before him, and he was flesh of her flesh, life of her life, and had inherited her "instinct" (as we call hereditary memory, in order to avoid the trouble of finding out what it is, and how it comes).'

Yet, in *Evolution, Old and New,* Butler (1879: 315-324) had already noted an ascription (1831) by Patrick Matthew, the Scottish landowner acknowledged by Darwin as having proposed evolution by natural selection decades earlier than himself (Schacter, 2001: 115):

> 'The continuation of family type [heredity], … is mental as well as corporeal, and is exemplified in many of the dispositions or instincts of particular races of men. These innate or continuous ideas or habits seem proportionally greater in the insect tribes, and in those especially of shorter revolution [lifespan]; and forming an abiding memory, may resolve much of the enigma of instinct, and the foreknowledge which these tribes have of what is necessary to completing their round of life, reducing this to knowledge or impressions and habits acquired by a long experience.'

In his 1838 *Notebook M* Charles Darwin wrote likewise: 'Now if memory of a tune and words can thus lie dormant, during a whole life time, quite unconsciously of it, surely memory from one generation to another, also without consciousness, as instincts are, is not so very wonderful' (Barrett et al., 1987: 521). In modern terminology, our minds do not begin as 'blank slates.' Some specific 'softwares' come 'factory-installed.' Butler was defended (anonymously) by Cambridge philosopher James Ward (1884), who accused Romanes of improperly attacking, while at the same time silently adopting, Butler's arguments (Pauly, 1982).

### Hering's essay approved by Bateson's mentor

Although Hering's work was mentioned in a *Nature* review of an essay by Haeckel (Lankester, 1876), it seemed to have little impact. Writing to Romanes in 1876, Darwin thought the essay 'clever and striking' and noted the 'stress on inheritance being a form of unconscious memory,' but was otherwise dismissive (Schwartz, 2010: 138-151). However, a Johns Hopkins University professor, William K. Brooks (1848-1908), quoted Haeckel extensively. There was 'a profound truth' in the view of Hering that 'Heredity is Memory,' (Brooks, 1883: 33-41):

> 'It is plain that this power [to adjust to the environment] depends upon experience, but experience depends upon "memory." So we may state, with truth, that in a certain sense, life is memory. … We know memory, however, only in connection with organization, and if it is true that heredity, the power of an organism to reproduce its like, is simply the memory, by the ovum, of the experience of its ancestors, we must believe that there exists in the ovum an organization of some kind to correspond to each of these past experiences.'

In 1883 and 1884 Brooks was visited by a young Cambridge zoologist, William Bateson (1861-1926), who in 1886 would be much impressed by a paper of Romanes on the origin of species (Cock and Forsdyke, 2008: 131-132). In 1900 Bateson would also be much impressed by an earlier paper of Gregor Mendel that he brought to the attention of the English-speaking world (Cock and Forsdyke, 2008: 198-220). Echoing Brooks, Bateson refered to genetic factors, not as conveying 'information for,' but as possessing 'the power to produce,' parental characters in their children (Cock and Forsdyke, 2008: 325). And, echoing Romanes (1893), in an address in 1914 (Cock and Forsdyke, 2008: 410-411) he supposed that:

'The elements upon which … [parental characters] depend … [are not] in any simple or literal sense material particles. I suspect rather that their properties depend on some phenomenon of arrangement. … By the re-arrangement of a very moderate number of things we soon reach a number of possibilities practically infinite.'

### *Richard Semon*

The Hering/Butler viewpoint, and Butler's extensive elaborations thereof, were adopted in Germany by Richard Semon (1859-1918), who coined a complex new terminology. But Semon considered Butler's work 'rather a retrogression than an advance' (Schacter, 2001: 122). Indeed, Francis Darwin praised Semon, while according Butler perfunctory acknowledgment (Cock and Forsdyke, 2008: 544). At the Darwin Centenary celebration, Bateson (1909) praised Butler, but concerning the 'Mneme-theory of Semon,' the latter's mentor, Haeckel (1909), was ecstatic:

'The most important advance that evolution has made since Darwin and the most valuable amplification of his theory of selection is … the work of Richard Semon … . He offers a psychological explanation of the facts of heredity by reducing them to a process of (unconscious) memory. The physiologist Ewald Hering had shown in 1870 that memory [stored information] must be regarded as a general function of organic matter, and that we are quite unable to explain … reproduction and inheritance, unless we admit this unconscious memory.'

Others, no less ecstatically, later recalled the appearance of *Die Mneme* as 'a comet, glowing and trailing a long tail' (Goldschmidt, 1956). Writing in German in 1925, Schrödinger ranked Semon with Spinoza and Schopenhauer, as one with whose writings he had been 'deeply involved' in his youth (Schrödinger, 1964: viii). However, by this time Butler's work had won some recognition in the English-speaking world and Semon was accused of improperly attacking, while at the same time silently adopting, Butler's arguments (Cock and Forsdyke, 2008: 549-553).

Sigmund Freud's writings on the unconscious mind were then gaining attention, but he acknowledged a debt neither to Butler (Parkin-Gounelas, 2007), nor to Semon. In 1884 Hering had invited the young Freud to become his assistant (Freud 1984: 211-212). In a book – *The Unconscious - An Introduction to Freudian Psychology* – philosopher Israel Levine (1923: 39-42) portrayed Butler as representative of 'Pre-Freudian views of the Unconscious,' linking him to Leibniz (1646-1716). However, in Anna Freud's translation into German of Levine's book, Freud himself chose to translate the Butler section, adding a footnote: 'German readers, familiar with this lecture of Hering's and regarding it as a masterpiece, would not, of course, be inclined to bring into the foreground the considerations based on it by Butler.' At a 1906 meeting of the Viennese Psychoanalytic Society, Freud had been critical of Semon's *Die Mneme* (Otis. 1994: 187-188):

> '[I] learned from Semon's book only that the Greek word for memory is *mneme*. The book is characteristic of those pseudo scientists who imitate exactitude merely by operating with numbers and concepts, and then feel they have accomplished something. Hering's idea, which is Semon's point of departure, is ingenious and subtle, but only the opposite can be said of Semon's work. Only he who has new things to say is entitled to coin new terms.'

But, while Butler lies buried as 'a classical crank,' Semon lives on (Schacter, 2001: 196)! Happily, the Clifford/Romanes version of ToM, as developed by James Baldwin (1861-1934) and Jean Piaget (1896-1980), also lives on (Obiols and Berrios, 2009; Frith, 2012), but seldom in the informational sense dealt with here and elsewhere (Tononi and Koch, 2014).[2]

## 4. Modern perspectives

### *Hereditary-information-flow*

For present purposes our understanding, in exquisite molecular detail, how Clifford's 'matter in motion' corresponds to the transfer of information through the generations, is largely irrelevant. But it does help us clarify the notion of information *as an abstraction*. At face value, a DNA sequence appears as a random sequence of its four base 'symbols' or 'letters,' – A, C, G and T. For example: TTCAGCCTCGTGGGGGACAAG. This sequence can be nothing more than what you see. So the molecule is but a vehicle for the bases of which it is composed. But it also, by virtue of the particular order and type of its base letters, *contains* information, just as the sentence you are now reading *contains* information. This abstract entity – the 'message' *within* a sequence – is *extracted* from the DNA by the cellular transcribing apparatus which assembles a corresponding RNA sequence without destroying the originating DNA (i.e. the latter's information is copied). The information *within* the resulting 'messenger RNA' is then *extracted* by the cellular decoding apparatus, which uses it to assemble a sequence of amino acid 'letters' – say FSPVGDK. Here the first 'letter' is F, but it could have been any one of the other nineteen amino acids. This is a part of a protein, and is encoded by the above DNA sequence and its messenger RNA copy. Thus, *information flows from medium to medium*, and if the flow is faithful, we can say that each medium has accurately received, preserved, and transmitted, it.

A protein folds into a structure determined by the order and type of its amino acid letters. It can then play some specific role. It may 'tell' a cell to accelerate a certain chemical process, or to build more membrane. Thus, in less discrete (analog) form, the information in the original message flows onward (Noll, 2003). While the details need not detain us, we can clearly recognize the initial flow of digital *coded information* from DNA to RNA and to protein, with each step of this flow being *entirely* dependent on the chemistry of, and energy provided by, the cell. *There is no step wherein the information is free of a medium*. But the information itself is an abstraction. The information is not the medium, but it is medium-based. In terminology that antedates modern biochemistry (Clifford, 1878: 278), the 'motion of matter' brings hereditary information to serve our practical, day-to-day, needs.

### Permanency and independence of 'selfish' information

Apart from this, hereditary information is preserved and passed through the generations. The medium – parental DNA – is replicated to produce DNA-bearing gametes. Here information flows from one DNA molecule to a structurally identical DNA molecule. The parents with the DNA they contain may then perish. Each new generation is defined by the information that is passed to it. Segments of DNA encoding information for cat characters, confer catness on offspring.

Such segments, of which the best understood are the genes, can be seen as independent entities. The essence of a gene is its information – its informational message – not the transient medium that contains it. This immortality of genes, which are preserved and transported by the mortal organisms that host them, was noted by Wilhelm Johanssen in 1923 (Cock and Forsdyke, 2008: 504-507), and by Hans Kalmus (1950), who remarked:

> 'A gene … is a message, which can survive the death of the individual and can thus be received repeatedly by several organisms of different [successive] generations. A gene may reproduce itself faithfully and in fact we do not know of any gene which can survive without doing so. … The permanency of [brain] memory as popularly understood has often been stressed – "the elephant never forgets" – but it is certainly surpassed by the permanency of the genes, which carry their messages through the generations.'

Here, of course, it is the *medium* – DNA – that 'may reproduce itself faithfully,' but the abstract informational message it contains does not just come along for the ride. The presence of the *proper* informational message ensures that DNA reproduction will be sustained. Genic messages endowing offspring with more advantageous characters will ensure, not only the survival of their hosts, but also their own survival, and thus, the future replication of the media that happen to contain them (Forsdyke, 2011b).

### *Macrocosmic information*

Kalmus's theme was taken up by Williams (1966: 22-25) and, more popularly, by Richard Dawkins (1976: 205-206),[3] who considered such 'selfish' genes as examples of universal 'replicators' that, along the lines of Romanes (1886: 49-50), could serve as models for macrocosmic forms of information:

> 'What … is so special about genes? The answer is that they are replicators. The laws of physics are supposed to be true all over the accessible universe. Are there any principles of biology which are likely to have similar universal validity? When astronauts voyage to distant planets and look for life, they can expect to find creatures too strange and unearthly for us to imagine. But is there anything which must be true of all life, wherever it is found, and whatever the basis of its chemistry? If forms of life exist whose chemistry is based on silicon rather than carbon, or ammonia rather than water, if creatures are discovered which boil to death at -100 degrees centigrade, if a form of life is found which is not based on chemistry at all, but on electronic reverberating circuits, will there still be any general principle which is true of all life? … I would put my money on one fundamental principle. This is the law that all life evolves by the differential survival of replicating entities. The gene … happens to be the replicating entity which prevails on our own planet.'

Thus, matter in motion would have varied until it happened to acquire both the ability to self-replicate, and the stability to sustain a chain-reaction, so allowing production of more of its kind. There was random variation in the material world until a form arose that 'made sense' informationally, such that differential survival was possible. The material world would have preceded the informational world. Natural selection's culling of these randomly generated different informational forms would have followed.

### *The limits of DNA*

Darwin came close to modern gene concepts when postulating fundamental units named 'gemmules' (Forsdyke 2001: 151), and Butler (1878: 187) supposed 'a memory to "run" each gemmule.' In 1923 the words 'memory' and 'hereditary continuity' entered Johannsen's discussion of genes (Cock and Forsdyke, 2008: 504-507), but 'memory' (i.e. stored information) did not acquired regular usage in this context until the 'information revolution' of the 1940s (Kalmus, 1950; Noll, 2003). However, the cybernetics pioneer Norbert Wiener (1950: 102, 182) went no further than noting:

'The biological individuality of an organism seems to lie in a certain continuity of process, and in the memory by the organism of the effects of its past development. This appears to hold also of its mental development. In terms of the computing machine, the individuality of mind lies in the retention of earlier tapings and memories, and in its continued
development along lines already laid out.'

We now know that a DNA sequence can simultaneously contain many, sometimes
overlapping, informational messages. These can convey both genic and non-genic information
that, given limited channel capacity, must compete with each other for genome space (Shadrin
and Parkhomchuk, 2014; Battail, 2014). So far, there is no evidence that DNA can sustain the
transfer of instinctual information between generations (Forsdyke, 2011a: 3-26). Thus the
informational bases of nervous systems (mental memory) and genomes (genetic memory)
differ (Forsdyke, 2006, 2009).

### The mental realm

Romanes and Butler were sons of clergymen and went to Cambridge with the intention of
taking holy orders. Romanes switched to the biological sciences at an early stage. Butler
completed the Classical Tripos and worked in a London parish in 1858 to prepare for ordination.
However, increasingly doubting earlier religious teachings – such as are described in *The Fair
Haven* (Butler, 1873a) and *The Way of All Flesh* (Butler, 1903) – he switched to practical biology,
reading Darwin's *Origin of Species* and Justus Liebig's *Agricultural Chemistry* while profitably
sheep farming in New Zealand (Forsdyke, 2006).

Confronted by dual variables – matter/motion and mind – Romanes pondered whether mind
('Spiritualism') was 'the ultimate Reality' that anteceded matter and motion ('Materialism'), or
vice versa. He concluded by equating them (Romanes, 1886). This non-dualist 'theory of
Monism' is described in his *Mind, Motion and Monism*. Thus, 'mental phenomena and physical
phenomena, although apparently diverse, are really identical' (Romanes, 1895: 83). Yet, his
other poetic, philosophical and religious works (Romanes, 1889, 1895; Turner, 1974: 134-163;
Pleins, 2014) suggest that monism did not satisfy his deep theological doubts.

A century later, Williams (1992: 3-4) also had doubts. He advised 'an open mind … especially
for neural phenomena. … I have no inclination to deny the mental realm or belittle its
philosophical importance. I am inclined merely to delete it from biological explanation.' Thus,

he declared himself not an atheist, but preferred to define atheism out of existence (Williams, 1997: 152-153): 'Whatever entity or complex of entities is responsible for the universe being as we find it, rather than some other way or not there at all, can be called *God*.'

Romanes (1882: 881-882) had distinguished two processes, those 'taking place in a something which is without extension or physical properties of any kind, and the other taking place in a something which … we recognize as having extension and other antithetical properties which we together class as physical.' Likewise, Williams (1997: 164-166) categorized three distinct 'domains' for processes: the material, the informational ('codical'), and the mental. A segment of DNA was in the material domain. The abstract potentially immortal gene that it might convey was in the informational domain. The mental domain was, somehow, separate from these.

As example, Williams gave the word 'guilt' that either 'can be a feeling (mental) about oneself or a public judgment (codical), perhaps as a result of a jury trial.' However, mental guilt can be seen as a mentalese-information-flow within the accused that may, or may not, be externalized in speech or text. Likewise, judicial guilt can be seen as mentalese-information-flow within a judge – an opinion (thought) that becomes action when it is externalized as a spoken or textual verdict. It seems unnecessary to disassociate informational and mental domains merely because one information-flow tends to remain internalized as a 'feeling,' whereas the other is converted into action. This was essentially the point made by Butler in his colourful 1887 lecture 'On the genesis of feeling' (Butler, 1923: 193): 'But reflection will show that opinion and feeling row in the same boat, and that as we form opinion so also we form the feeling on which the opinion is based and from which it springs – feeling being only opinion writ small.' He held that a 'grammar of feeling' had evolved early, in much the same way that a grammar of language had evolved (Butler, 1923: 205-206):

'Even our most well defined and hereditary instinctive feelings were in the outset formed … by long and arduous development of an originally conventional arrangement of sensation and perception – symbols – caught hold of in the first instance as the only things we could grasp, and applied with as little rhyme or reason as children learning to speak apply the strange and uncouth sounds which are all they can then utter. … When I say "we," I mean, of course, our remote invertebrate ancestors, … [who] attached some of the few and vague sensations, which were all they could then command, to such motions of outside things as echoed within themselves, and used them to feel; hence, presently, by repetition of the same process, they invented a second conventional arrangement of ideas – symbols – to think the things with which they connected them, so as to docket them and recognize them with greater force, certainty, and clearness; much as we use words to help us docket and grasp our ideas and feelings, or written characters to help us docket and grasp our words. Having once established the connection they stuck to it, not probably without much wrangling … , nor yet without the growing up of many different languages, as it were, of feeling, some of which doubtless still prevail among the … lower existing forms of life. … On this, the development of a rude grammar of feeling, and of the nervous mechanism whereby its rules could be alone formulated and carried into practice, were a mere question of time … .'

By virtue of his firm grasp of the information concept, Butler was better able to cast aside religious trappings to arrive at an essentially similar viewpoint as Romanes, while avoiding much of his theological angst (Forsdyke, 2006, 2009, 2011a: 7-26). In *Unconscious Memory*, Butler quoted Hering with approval (Butler, 1880: 102-103):

'The materialist regards consciousness [mind, spirit] as a product or result of matter, while the idealist holds matter to be a result of consciousness, and a third maintains that matter and spirit are identical; with all this the physiologist, as such, has nothing whatsoever to do; his sole concern is with the fact that matter and consciousness are functions one of the other.' [Thus, mind and matter are] 'two variables … so dependent upon one another in the changes they undergo in accordance with fixed laws that a change in either involves simultaneous and corresponding changes in the other.'

### *Failure to communicate*

However, as with many others who came to study Butler's writings, the historian Frank Turner (1974: 164-200) did not see memory in informational terms. Despite Butler's satirical style that many recognized, Turner took literally Butler's self-proclaimed amateur status. Butler's 'patchwork' writings were considered to lack 'philosophical rigor,' and some, such as *The Fair Haven,* were declared 'a total failure.' Turner concluded that 'there was nothing in Butler's writings to suggest that he believed the presence behind reality itself had any essence … .' To Turner, this non-material lack of essence meant a 'spiritual,' not 'informational,' abstraction.

If Butler failed in any way, it is perhaps that his fair-minded personal ToM led him to look horizontally across to, rather than down on, those who might be his peers. In short, he did not realize just how far ahead his own thinking was. The truculence he so often encountered was seen as willful, rather than as an expression of bewilderment. When *The Fair Haven* was attacked, Butler (1873b) responded: 'The main purpose of the book was not argumentative, but satirical. … A writer cannot write for everyone; he must assume a certain amount of apprehensiveness on the part of his readers and is justified in leaving children and stupid people out of his calculations.'

### Outsourcing long term memory

Romanes had gone beyond Clifford in proposing that extracorporeal mindstuff could become organized, but in 1886 he did not consider how this proposed 'world eject,' perhaps with its own subjectivity, might relate to the apparently localized subjectivity of an individual brain. Conflicting with the view that 'the phenomena of mind are invariable in their association with cerebral structures' (Romanes, 1882: 880), the brain as 'the exclusive seat of mind' is today challenged by considerations of the massive storage space needed both by human brains for long-term memory, and by gametes for the transfer of instinctual information into the long-term memory of the embryos they generate (Forsdyke, 2009). Indeed, recent reports hint that such information might be stored in, and retrieved from, an extracorporeal source, by a process analogous to 'cloud computing.' This source could correspond, in part, to Romanes' world eject. Evidence that part of our minds might not be cranially localized, challenges no less a philosopher than Ludwig Wittgenstein (1969: 18e) who, in *On Certainty*, pondered: 'Now would it be correct to say: So far no one has opened my skull in order to see whether there is a brain inside; but everything speaks for, and nothing against, its being what they would find there?'

Extreme examples of extracorporeal localization could be rare individuals whose childhood hydrocephaly was thought to have been successfully treated and are functionally normal – one even obtaining a first class degree in mathematics. Yet they are found to have *only 5%* the normal volume of cerebral tissue (Lewin, 1980). While initially disbelieved, there are now two further reports of relatively normal individuals with such minute brain tissue volumes. While, as noted by Makorek (2012) and Forsdyke (2014, 2015), these cases stretch the credibility of brain 'plasticity' explanations (the 5% able to become functionally close to 100%), or 'redundancy' explanations (we normally use 5% and the remaining 95% is superfluous), many neuroscientists still disbelieve. Extraordinary claims require extraordinary evidence, and neurones are known to be remarkably stress-resistant (Ding et al., 2001). But there are growing indications that the hallowed foundations of modern neuroscience are not as stable as once supposed (Firestein, 2012; Satel and Lilienfeld, 2013).

And that complex 'factory-installed' behaviors (instincts) can be accounted for on the basis of the information content of gametes alone is also hard to believe. Outsourcing to a parallel extracorporeal racial memory – a 'collective unconscious' (Jung, 1959) – that could be tapped into during neural development, is no less hard to believe. Yet, failing the emergence of other explanations, both these arguments-from-incredulity deserve a place at the table of responsible neuroscientific discourse. While the critical 'crossroads neuroscience has reached' is gaining recognition, a multimillion dollar 'brain initiative' (Yuste and Church, 2014) may err if too narrowly focused on that wonderful organ.

## Conclusions

Inspired by Darwin and well versed in the classic languages, mathematics, philosophy and theology, Victorians such as Clifford, Romanes, and Butler, seem to have been better able to think broadly on biological problems – see the big picture – than many, detail-laden, later scientists and philosophers. The information concept entered biology with Hering and Butler, and passed by way of Semon and Schrödinger to illuminate the emergent discipline of molecular biology (1945-1966). Romanes was able to extrapolate Clifford's ToM hypothesis from individual humans to their societies – a 'stepping stone' to mind in the universe. The relationship, if any, of this higher order of subjectivity ('world eject') to that of individuals, was clouded by his persisting theological concerns. However, recent reports of normal memory, and even advanced intellect, in rare individuals with greatly reduced brain volume, would be consistent with an accessible extracorporeal long-term memory that might be part of such a higher subjectivity. Whether of external or internal origin, mentalese-information-flows ('thoughts') interact to generate the 'meanings' that, when we are awake, exist as information-flows in our conscious and unconscious minds and, when we are asleep, exist as information-flows in our unconscious minds (Majorek, 2012; Mashoura and Alkire, 2013).

## Acknowledgements

Gérard Battail and Chris Frith provided constructive comments on the manuscript.

## Notes

1. Religious texts have so shaped our language over millennia that we now use words, such as 'goodbye' (presumably 'God be with you'), with little thought for religious implications. Thus, it does not follow, since we borrow words from the domain deemed theological, that we necessarily subscribe to the beliefs of that domain. By the same token, just as politicians act prudently by being seen to practice religious faith, so daily prayers for family and servants was the norm in some Victorian households (Butler, 1903), where it would have been important to attract and retain dutiful servants, and to 'properly' educate children (Schwartz, 2010: 25-27).

2. Aspects of this paper overlap with the 'Integrated Information Theory' (IIT) of Tononi and Koch (2014), who equate consciousness with subjectivity – a definition similar to that of Majorek (2012). However, the scope of this subjectivity is more limited than proposed here. There is no Zeitgeist. Thus Tononi and Koch 'consider two people talking: within each brain, there will be a main complex – a set of neurons that form a maximally irreducible whole with definite borders … . Now let the two speak together. …  According to IIT, there should indeed be two separate experiences, but no superordinate conscious entity that is the union of the two.' For present purposes, two people communicating is the same as two parts of an individual brain communicating. In both cases, information flows can interact to give an output greater than the sum of the whole.

3. While a putative path to the selfish gene idea is described here and elsewhere (Forsdyke, 2011b), evidence for *direct* antecedent influences is unclear, with the exception of Williams who is cited by Dawkins. But ideas usually arrive first, just as ideas. The words ideas subsequently 'fly with' (Butler 1904: 187) can aid in their comprehension and marketing. Different words *compete* for this role. The words Johannsen, Kalmus, and Williams each assigned to the *idea* of the selfish gene, lost to the 'selfish gene' of Dawkins (1976).

---